\pdfoutput=1 

\documentclass[11pt]{article}
\usepackage[margin=3cm]{geometry}
\usepackage{amsfonts,amsmath,amssymb,amsthm} 
\usepackage{times}
\usepackage{enumerate}
\usepackage{subcaption}
\usepackage{verbatim}
\usepackage[bookmarks=false,colorlinks=true,urlcolor=blue]{hyperref}
\usepackage[affil-sl]{authblk}  
\usepackage{url}
\usepackage[normalem]{ulem} 
\usepackage[inline, shortlabels]{enumitem}   
\usepackage{cite}  
\usepackage{graphicx}
\usepackage{wrapfig}

\title{Retrospective Observation of Ball Lightning}
\author{Neil J. Gunther}
\affil{\small Performance Dynamics Research, Castro Valley, California, USA }
\date{}

\begin{document}
\maketitle
\thispagestyle{empty}

\begin{abstract}
Ball lightning is a very rarely observed phenomenon that makes it  extremely difficult to characterize scientifically. 
Eye-witness accounts of ball lightning that describe many features in common have been variously recorded for centuries. 
This note is an account of my observation of ball lightning almost five decades ago. 
Although it is mostly a compendium of recalled impressions (no photos or videos), 
it so happens that certain corroborating information has since become available on the Internet and  
supports the conclusion that I did indeed observe ball lightning. 
This note also helps to explain why obtaining hard evidence is even more difficult than is generally assumed. 
Paradoxically, the inordinate delay between the original observation and this documentation  
may actually have enhanced its scientific merit. 
\end{abstract}

\section{Introduction}   \label{sec:intro}
Ball lightning is a rare and mysterious natural phenomenon~\cite{wikiBL} for which 
eye-witness accounts have been variously recorded over centuries~\cite{book1949,boerner2007,QLD2011,keul2021,medieval2022}.
In a BBC radio interview~\cite{bbc},  Prof. Karl Stephan (Texas State University)  explained how his team is  
using an online survey~\cite{survey} to seek detailed eye-witness accounts, preferably including smartphone 
images and videos, that could assist in better characterizing the physics of ball-lightning~\cite{eos2021,keul2021}. 
Some common attributes, mentioned in the interview, include:
\begin{itemize}
\item {\bf Size} of about 20--25 cm 
\item {\bf Brightness} of an incandescent light bulb 
\item {\bf Lifetime} on the order seconds 
\item {\bf Movement} that is not affected by wind or gravity 
\item {\bf Locomotion} that appears independent of any obvious forces
\item {\bf Electrical} in nature: as evidenced by \href{https://en.wikipedia.org/wiki/Lichtenberg_figure}{Lichtenberg patterns} and damage to electrical equipment
\end{itemize}
An Appendix to this note contains additional characteristics historically identified with ball lightning. 
It could also be that different creation mechanisms produce different variants of ball lightning. 
The attributes of the ball-lightning event described here are discussed in Section~\ref{sec:summary}.

Whereas typical lightning, e.g., cloud-to-ground (CG) lightning~\cite{nwsCG}, is regularly and universally observed, 
the much rarer occurrence of ball lightning, combined with a lack of direct and repeatable measurement\cite{boerner2007,QLD2011}, 
makes its physics~\cite{usaf2003,sturrock2016,sturrock2017,boerner2018,funaro2018,georgia2019} even more speculative than mechanisms for common lightning.  
Can the citizenry network of eyeballs and smartphones help to improve this situation?

In gathering citizen responses, the Texas State University online questionnaire  accepts text input only. 
The space provided for answering each question is rather restricted but flexible in that it can be 
revisited and edited---unlike Twitter, for example. 
Any images, videos or other supporting media need to be delivered via ancillary email. 
Since I believe I observed a ball lightning event  (BLE hereafter), I filled out the Texas State survey.  

As a result of completing that  questionnaire,  however, 
I started to have some doubts about certain of my responses regarding the BLE that I witnessed decades ago. 
For example, {\em precisely} what year did witness the BLE? 
My best recollection was some time in the early 1970s. (See Section~\ref{sec:year} for the final resolution.)  
Although I have carried certain vivid impressions with me over the years,\footnote{It is now well-established scientifically that people's memories and impressions are notoriously inaccurate, despite how much personal stock they put in them.} 
I started to realize that, if pressed (as I was in answering the questionnaire),  
I had no way of verifying the precision of some of my recollections. 
Then, it occurred to me that it might be possible to find supporting evidence using various Internet 
resources, such as Google Maps and meteorological databases. 
Thus, the following personal account is much more detailed than was feasible using the short online questionnaire~\cite{survey} .

\section{The Event}  \label{sec:event}
Some time during the late spring or early summer of 1973 a severe electrical storm occurred in and around the city of 
Melbourne, Australia. Specific details, such as date, time and GPS coordinates, are given in Section~\ref{sec:summary}. 
The storm started in the late afternoon or early evening and was accompanied by intermittent rain. 
The atmospheric humidity was high.  CG lightning activity was unusually intense~\cite{boerner2018}. 

\begin{figure}[!ht]
    \begin{subfigure}[t]{0.48\textwidth}
        \includegraphics[scale=0.305]{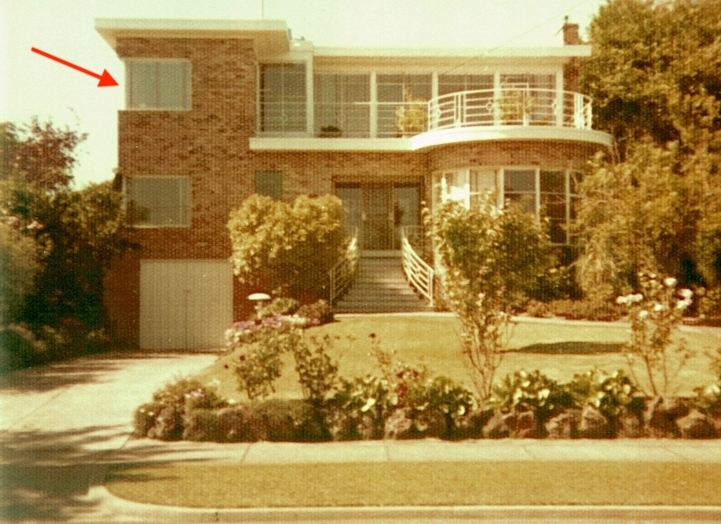}
        \caption{Observation point located in the upper-story corner windows ({\em arrow}) 
        as it would have appeared in the summer of 1973.}    \label{fig:bedroom}
    \end{subfigure}
    \hspace{0.20in}
    \begin{subfigure}[t]{0.48\textwidth}
        \includegraphics[scale=0.305]{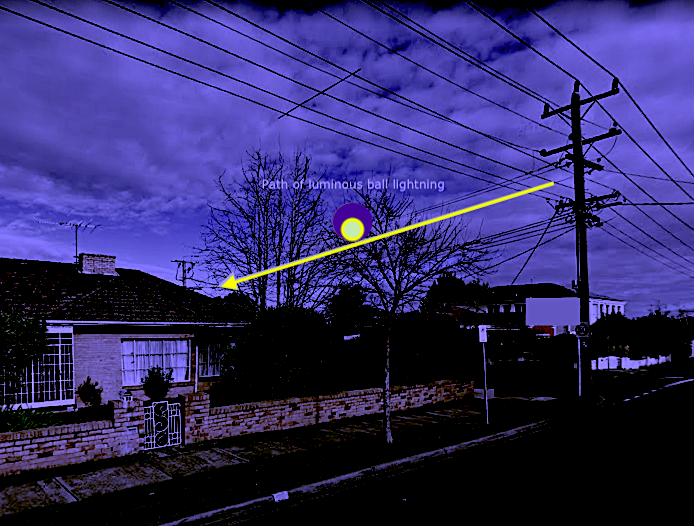}
        \caption{Location of lightning-ball and its trajectory between power lines 
         based on image sampled in the winter of 2019.}      \label{fig:fireball}
    \end{subfigure}
    \caption{Observation point and location of the ball-lighting  event.}
\end{figure}

This unusual intensity caused both my father (an electrical engineer who was always excited by lightning) and myself to watch the storm from my bedroom windows on the uppermost floor of our house in Fig.~\ref{fig:bedroom}.  
That would have placed us both at about 25 ft (7.5 m) above street level. 
According to Google Maps, the ground distance from that corner of our house to the power lines in  Fig.~\ref{fig:fireball} is about 120 ft (36.5 m). 

\begin{figure}[t]
 \centering
    \begin{subfigure}[t]{0.475\textwidth}
     \centering
        \includegraphics[scale=0.2375]{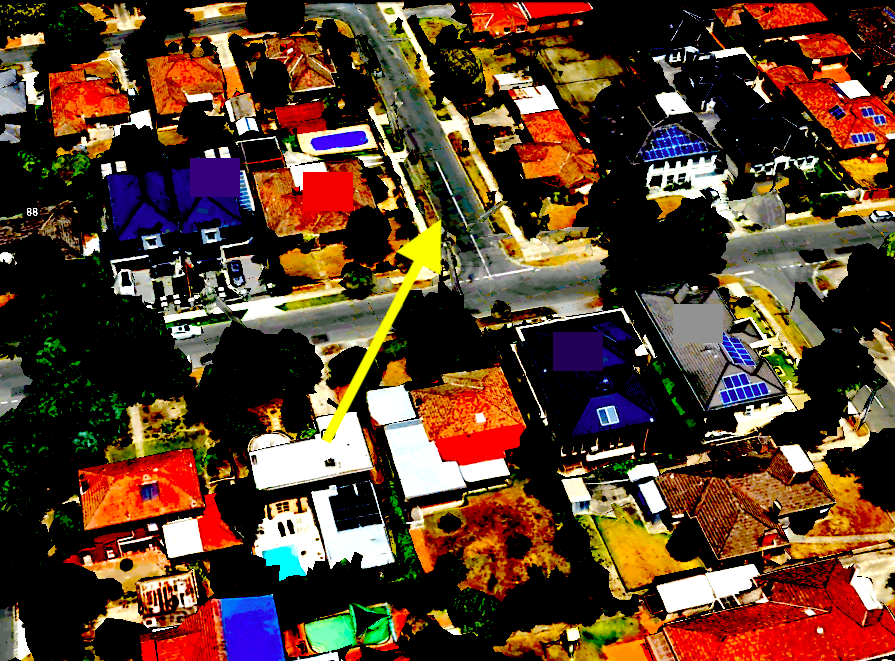}
        \caption{View from well above the observation point to the power lines running down the side street.}     \label{fig:3d-house}
    \end{subfigure}
    \hspace{0.15in}
    \begin{subfigure}[t]{0.475\textwidth}
     \centering
        \includegraphics[scale=0.24]{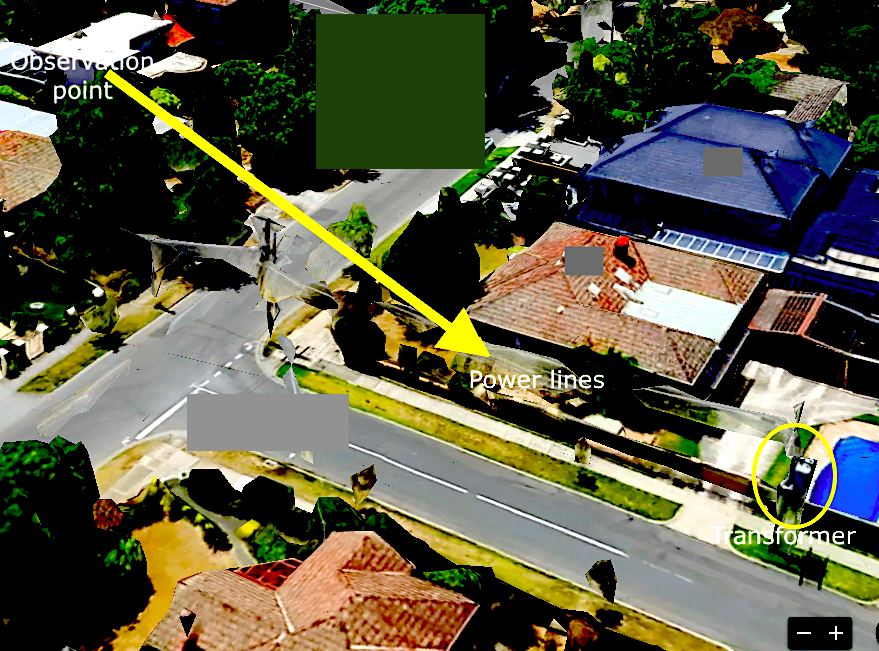}
        \caption{Relative positions of the observation point and the BLE as seen from the side street.}    \label{fig:3d-sidest}
    \end{subfigure}
    \caption{Elevation views of the observation point above the BLE location.}   \label{fig:3d-goog}
\end{figure}

Our house faced logical south and was located approximately 6 miles (10 km) due west-northwest of Melbourne city center.
The panoramic view from my bedroom was slightly greater than a quadrant spanning from due south to due west. 
The radial viewing distance differed greatly depending on various undulations in the landscape, 
tree density, and clustering of suburban buildings. 
The longest line of sight, on a clear day, would have been due west for approximately 10 miles (16 km).

After watching the CG lightning for some time, it had become dark.\footnote{Darkness was accentuated and accelerated by the eastward movement of the storm clouds from the Melbourne city area. The prevailing winds at that time of year are westerly. } 
I happened to be looking in the direction of the power lines (i.e., approximately south south-west)  
when I noticed a light moving at constant speed in the wet trees at approximately the height of the power lines 
shown in Fig.~\ref{fig:fireball}. 
This turned out to be ball lightning but, that was not immediately clear at the time, despite the light giving off 
sporadic sparks. The concomitant sparking and mottling,  discussed in Section~\ref{sec:sparks}, did not register immediately as being significant. 

\begin{figure}[htbp]
\centering
\includegraphics[scale=0.30]{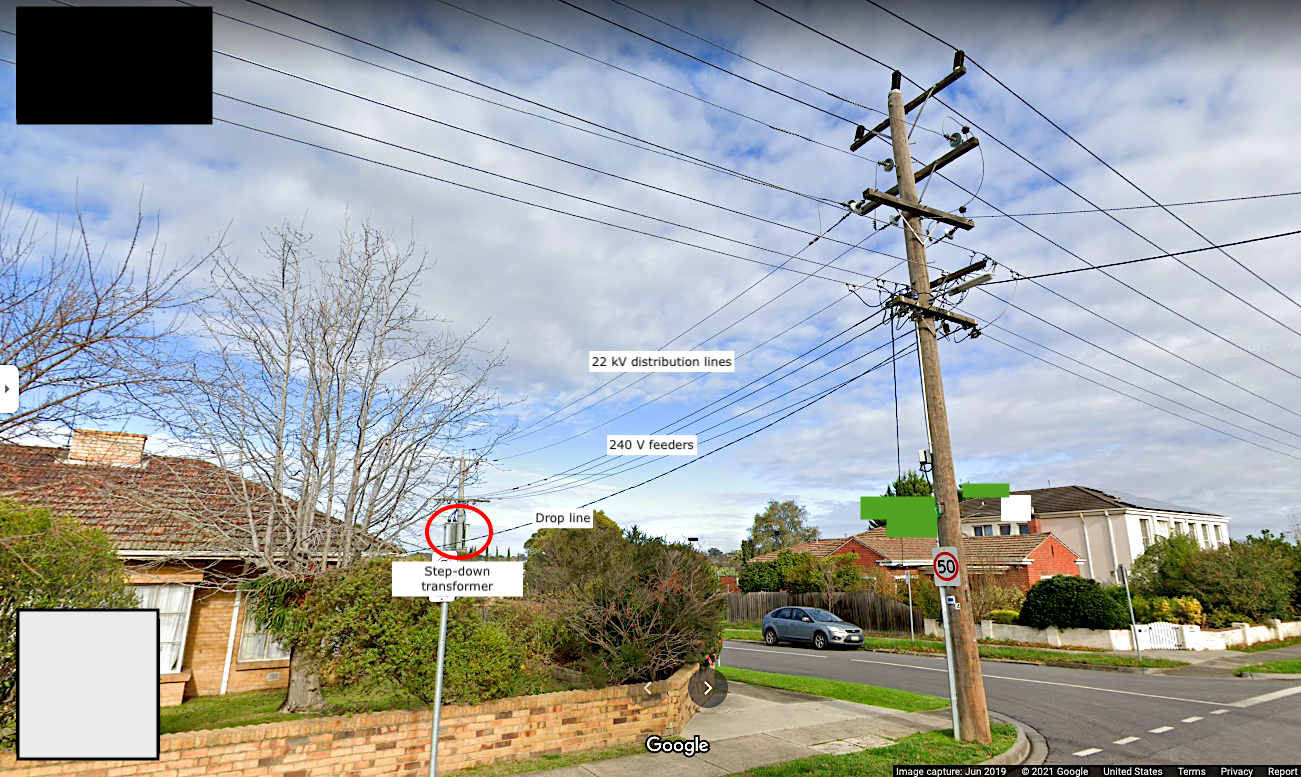}
\caption{Above-ground suburban power distribution tiers  showing 22 kVAC high-tension lines, step-down transformer, and domestic 240 VAC drop lines. (cf. Fig.~\ref{fig:fireball})}   \label{fig:wires}
\end{figure}

Fig.~\ref{fig:fireball} is a {\em simulated} evening view of the street corner forty-six years after the BLE (Google Street View 2019). 
It is based on the corresponding daytime image  in Fig.~\ref{fig:wires}. 
Since that image was captured during the winter of 2019, the relevant deciduous trees are leafless. 
Remarkably, the above-ground power lines and the trees are the same as they were 5 decades ago:  
the main difference being that it was dark during the BLE in 1973.   
Moreover, the trees had leaves (being late late spring or early summer) that were wet and reflective from the intermittent rain during the storm.

The  light (depicted schematically as the pale green disk in Fig.~\ref{fig:fireball})  seemed to travel just above the lowest tier of power lines, which I estimate to be at a height of 15 ft (4.5 m).
The downward viewing angle to the BLE was therefore $\arctan[(25 - 15) / 120] \simeq 5^o$ or just below eye level. 
Using Google Earth (2017), 3D elevation perspectives are provided in Fig.~\ref{fig:3d-goog}. 
Many of the surrounding neighborhood trees have grown considerably since 1973. 
Shortly after noticing the pale round light, it disappeared behind the roof of the corner house. 
If our observation point had been another 25 ft (7.5 m) higher, as depicted in Fig.~\ref{fig:3d-goog}, we would likely have able to see the light continue down the side street for another 250 ft (70 m), or more, as it moved along the power lines.

Since the light appeared to be moving through the trees and had the intensity of reflected automobile headlights,  
my first impression was that the light was due to a vehicle turning into the side street---the mottled appearance of the light being due to reflections off the random individual surfaces of the wet leaves. 
The light was in view for about 5 seconds before it disappeared behind the roof of the corner house. 
If the length of the visible wires in Fig.~\ref{fig:fireball} is about 25 feet (7.5 m), the apparent speed of the moving disk would be about 5 ft/s or 4.4 mph (1.5 m/s), 
which matches the expected  angular sweep speed of a car turning that corner.  
Although a vehicle has two headlights, they converge to a single spot or area and would therefore appear as a single disk of light. 
See Sections~\ref{sec:headlights} and~\ref{sec:sparks} for an explanation of why this impression was incorrect. 

Seen from a distance, the luminous disk appeared to have a diameter close to that of a basketball, viz., about 1 ft (30 cm), 
and had a pale green glow. 
If the color had been different, e.g., purple or blue, it would have stood out as being obviously different from the green tree leaves and thus, would not have evoked the impression of vehicle headlights after all.

Figure~\ref{fig:wires}  provides a more detailed recent image (Google Street View, 2019) of the power distribution lines in Fig.~\ref{fig:fireball}. 
In suburban Melbourne, the uppermost tier 
\href{https://electronics.stackexchange.com/questions/356436/australian-suburban-power-distribution}{typically carries 22,000 VAC} from a local substation. 
A transformer steps this voltage down to 240 VAC feeders that connect power to each house via the lowest tier and individual drop lines~\cite{melpwr}.  
The ball of light moved horizontally down the side street, between the \mbox{22 kV} and \mbox{240 V} power lines, towards the 
step-down transformer.

My father and I both watched the thunderstorm until its intensity eventually abated. 
He, in particular, did not want to miss anything.
As a result, I can state categorically:
\begin{enumerate}
\item No power outage occurred in our neighborhood.  \label{item:outage} 
\item No nearby CG lightning strikes occurred    
within our field of view: the quadrant from south to west. There could have been unseen strikes in the other three quadrants. 
There certainly was thunder. \label{item:strikes} 
\item No other BLE observations were reported for this storm. 
And, being such a transient observation, it never occurred to us to report this BLE.  \label{item:unique} 
\end{enumerate}

\noindent
Point~\ref{item:outage} suggests that what I witnessed was not some kind of electrical arcing or short circuit in local the power distribution system. 
Point~\ref{item:strikes} would seem to support the hypothesis that CG strikes and the occurrence of 
 BLEs are not necessarily locally correlated in space~\cite{boerner2018}. 
Point~\ref{item:unique}, regarding reported meteorological observations, is discussed in Section~\ref{sec:year}.

It is ironic that my father---a much bigger fan of lightning storms than me---did not directly observe the BLE since he happened to be looking in a much more westerly direction. 
By the time I exclaimed that I thought I just seen ball lightning, the BLE was already out of view. 
On the other hand, he was quick to agree that it most likely was a BLE.

\section{Was It Ball Lightning?} 
How do I know it was really ball lightning?  The simple answer is, I don't. 
Without hard physical evidence, I cannot be one hundred percent certain---the bane of this entire subject. 
Nonetheless, there is now quite a bit of circumstantial evidence that supports the   
conclusion that the ``reflected headlights''  were, in fact, a BLE.

\subsection{Mysterious headlights}   \label{sec:headlights}
My initial impression that the luminous disk in the trees was reflected vehicle headlights simply does not hold up on geometrical grounds. As reported in Section~\ref{sec:event}, the  angular sweep speed of the light did match that of a vehicle turning the  corner in the dark.

For the disk to move in a southerly direction in Fig.~\ref{fig:fireball}, a car would have to be making a right turn into that street. Automobile headlights are generally mounted a couple of feet above the road and aimed {\em down} the road. 
A vehicle making a right turn (even allowing for British driving system used throughout Australia) would not produce any light at the height of the trees in Fig.~\ref{fig:fireball} or Fig.~\ref{fig:wires}. 
Moreover, the twin headlight-reflectors are required by law to be focused in such a way as to converge onto a single broad overlapping area, not a circular disk.  

Most importantly, no cars were observed near that street corner during the BLE. 
Indeed, it is that simple fact that quickly led me to realize the luminous disk was not due to vehicle headlights. 
Overall, the height of the luminous circular disk was inconsistent with car headlights at street level.  
The disk shape of the light was wrong for it to be due to car headlights, and  no car was observed in the vicinity during the BLE.

\subsection{Damaged tree}  \label{sec:tree}
Out of heightened curiosity, the following day, I walked down the side-street where the BLE had been observed  
to see if there was any physical evidence. 
Seeing nothing in that street, 
I continued beyond the end of the power lines and into a nearby park---a large park area located roughly half a mile southward from our observation point. 

\begin{figure}[t]
    \begin{subfigure}[t]{0.46\textwidth}
        \includegraphics[scale=0.227]{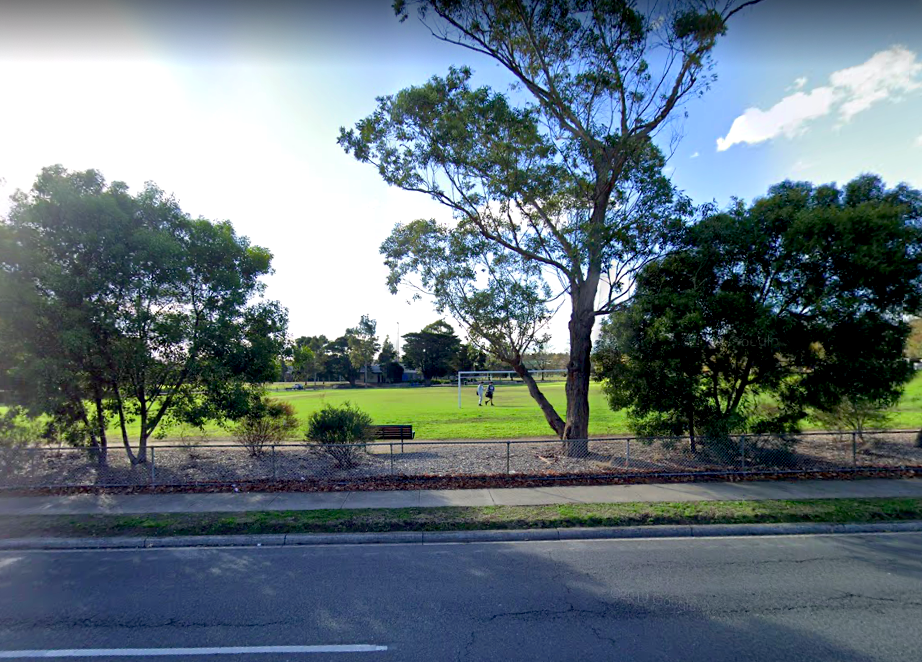}
        \caption{Looking north to the BLE origination point (Fig.~\ref{fig:fireball}) through the many trees in the park.}    \label{fig:trees}
    \end{subfigure}
    \hspace{0.250in}
    \begin{subfigure}[t]{0.46\textwidth}
         \includegraphics[scale=0.24]{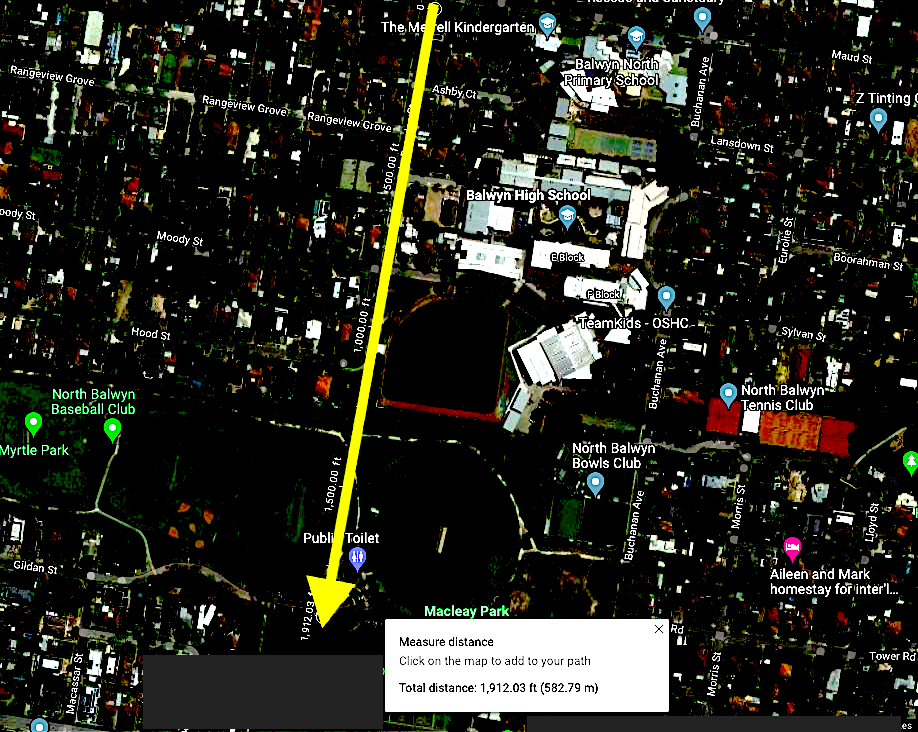}
        \caption{Trajectory (yellow arrow) from BLE origination to southern terminus area in the park.}     \label{fig:path}
    \end{subfigure}
 \caption{Plausible linear trajectory from BLE point of observation to termination in a local park.}   \label{fig:park}
\end{figure}

Within the park there were numerous trees, and that is still true today,
as can be seen in Fig.~\ref{fig:trees} via Google Street View, 2019.  Among them, 
I spotted a eucalyptus tree (or ``gum tree,'' in the local vernacular) with a large branch sheared at the trunk and exhibiting burn marks. 
The branch was still attached to the trunk by its bark but it was otherwise bent down toward the ground. 
The shearing point on the tree was roughly the same height as the power lines.  
No Lichtenberg patterns (cf. Section~\ref{sec:intro}) were observed in the tree bark.
Although the trees in Fig.~\ref{fig:park} were mostly hidden from our field of view in Fig.~\ref{fig:bedroom} by the rooftops 
in Fig.~\ref{fig:3d-goog}, no CG lightning strikes were observed in that termination area during the storm. 

More importantly, however, looking back from that tree along the side street with the power lines,  
there was a direct line of sight to the point of BLE creation as shown in Fig.~\ref{fig:path} via Google Maps, 2021. 
The estimated distance is about 2000 ft (600 m) or about one-third of a mile.
In other words, although there was no observation in this location, 
the lightning ball might have continued for several minutes~\cite[p.1438]{QLD2011} (assuming constant velocity) more or less horizontally along a nearly linear path beyond the point at 1250 ft (380 m) where the power lines are routed westward along a different side street. 

Power lines are not a necessary requirement  to support persistent, stable, BLE locomotion (cf. the characteristics  in Section~\ref{sec:intro}).  
On the other hand, in this case, the power lines might have played a very important role in the creation of the lightning ball,   
as well as its initial propagation---perhaps including {\em ballistic acceleration}. 

Whereas the observed horizontal motion between the power lines could be associated with parallel electric fields, 
the electric field beyond those wires would more likely be orthogonal to the surface of the ground and thereby 
introduce a vertical component to the motion. However, given that other effects such as, air currents and electrostatic reflection, could have been in play, any vertical motion of the lightning ball might have been oscillatory with the average height remaining close to its launch point from the power lines.   

Given the damaged tree and the absence of any local CG lightning strikes (point~\ref{item:strikes} in Section~\ref{sec:event}),  
it is hard to imagine what else could have produced such localized charring to the tree branch. 
The explosive termination of BLEs is well-documented~\cite{sturrock2016,keul2021,eos2021,funaro2018,georgia2019,wu2016}.

\begin{quote}
{\em On an oppressive day in Scotland in 1947 in which, however, there was no rain or thunder, a fireball was seen running along an outside electric wire. It struck a very large oak with a terrific explosion, shattering the tree to pieces.} 
---S. Singer, 1971~\cite{sturrock2017}
\end{quote}

\noindent
In this case, the damage was moderate and therefore not likely to have been reported.  
Eucalyptus trees contain highly flammable sap (gum) and the wood is typically very gnarly. 
The higher energy of a regular CG strike would be more likely to ignite the 
 sap and oil in the eucalyptus leaves---how many  bushfires start in Australia---possibly splitting the entire tree trunk.

\subsection{Sparks and mottling}  \label{sec:sparks}
The actual {\em ball} of light in Fig.~\ref{fig:fireball} had a phosphorescent, pale green color (near 555 nm), which may be due to 
 atmospheric oxygen being ionized inside the ball~\cite{QLD2011}.  

What I am less sure about today is, its non-uniform, mottled, appearance. My initial interpretation was that the mottling was due to 
headlights being reflected off individually oriented wet tree-leaves.  However, with the headlight explanation eliminated in Section~\ref{sec:headlights}, the mottling was presumably intrinsic to the lightning ball itself 
and might be associated with hydration of molecular ions~\cite{Turner2003}. 

Similarly, the sporadic sparks, alluded to in Section~\ref{sec:event}, were also initially misinterpreted as being the result of some kind of refractive effect related to the wet leaves.

\subsection{Horizontal motion}
Despite the incorrect initial impression (that the luminous disk was reflected headlights of a vehicle turning the 
street corner), its apparent horizontal motion can used to estimate the speed of the BLE.

A vehicle turning a corner, without first coming to a complete stop, can be estimated to be somewhere between 
3 to 5 mph or 4 to 5 ft/s (1 to 1.5 m/s). This apparent angular ``turning'' speed would appear {\em constant} over that short distance, 
in agreement with Section~\ref{sec:event}, 
In other words, the apparent horizontal velocity of the BLE could not have been more than a 
few feet per second, which is consistent with Section~\ref{sec:intro}. 

Curiously, that speed also happens to be close to the phase velocity of shallow water waves.  
Since air can be considered as a low-density fluid, it is as if  lightning  ball 
behaves like some kind of least-energy 
electric or electrochemical {\em bubble} in the atmosphere~\cite{wu2016,funaro2018,Turner2003}. 

In contradistinction to lab-created plasma bubbles~\cite{Planck}, 
no vertical motion, due to thermal convection~\cite{wu2016},   
was observed in the BLE reported here. The motion appeared to be entirely horizontal: certainly while it was between the power lines in Fig.~\ref{fig:fireball}.
Perhaps  the lightning ball is energetically excited  but internally is electrically {\em neutral}: 
analogous to superconducting~\cite{sturrock2016,sturrock2017} Cooper-pairs~\cite[\S 5.4]{wikiBL}. 
in which case a better question might be  
not what prevents the lightning ball from flying apart under Coulomb repulsion but, what prevents it from instantaneously collapsing?

\section{The Year}   \label{sec:year}
In the back of my mind, I always remembered the event as having occurred in the early 1970s. 
As I filled out the online questionairre~\cite{survey}, however, I became rather insecure about that memory and realized I did not have any way of corroborating the precise year, let alone the day.  
After almost 50 years since the event, it seemed hopeless to even narrow down the year.
The only things I was certain about were:
\begin{enumerate*}[label=(\roman*)]
\item I was either a teenager or in my early twenties, and 
\item my parents moved to a new house in 1975. 
\end{enumerate*} 

Then, I had a brainwave. Since the thunderstorm was severe, the Victorian branch of the Bureau of Meteorology (\href{https://twitter.com/BOM_Vic}{@BOM\_Vic})~\cite{bom} might have  a database of such occurrences and those data might be accessible online. 
As it turns out, not only does \href{https://twitter.com/BOM_Vic}{@BOM\_Vic} have records, they go as far back to the 18th century!
Surprisingly, for the period 1960--1975, the only severe lightning storms occurred in either 1973 and 1975. 
Since my parents sold the house in 1975, that makes 1973 the only possible year for my observation.

The Victorian Severe Storms database~\cite{bom} revealed that the thunderstorm occurred on October 31, 1973, which is late spring/early summer in the southern hemisphere. 
It seems to have started near Melbourne city center at 0530 UTC, which is 1530 AEST.  
As usual, the storm system would have gradually moved eastward. 
\href{https://twitter.com/BOM_Vic}{@BOM\_Vic}'s definition of {\em severe} refers to the fact that two people were killed by CG lightning and a boy drowned due to flash flooding.  
These incidents occurred as far away as 20 miles (31 km) east of the Melbourne city center. 
The BLE described here occurred about 6 miles (10 km) east of  downtown Melbourne.

\section{Image Capture}
It is important to underscore the great difficulty of capturing images of a rare event like a BLE, 
despite the ubiquity of modern smartphone cameras. 
In my case, even if I had possessed a smartphone in 1973,\footnote{As a student, I did not even own a camera until 1977.}    
it remains highly unlikely that I would have captured any image of the BLE.
The  problem is that when the BLE occurred,  I thought initially it was merely reflected  headlights and  
I would not have been motivated to capture any image. 
The chances would have been much higher if I had already been taking snapshots of the intense CG lightning---the reason we were watching the storm in the first place.

The chances would have been {\em maximized} if 
I happened to be recording {\em video}---multiple image frames---in the direction where the BLE occurred, at precisely that moment. 
Indeed, that is how a unique BLE spectrograph was captured in 2014~\cite{spectral2014}.
The following, somewhat whimsical,  space-time diagram explains why the probability of 
image capture is so small. 

\parbox{2.75in}{
Four world-lines are depicted schematically:
\begin{enumerate}[label=(\alph*)]
\item Lightning ball creation and propagation ({\em green arrow}) 
\item Human observer  ({\em red})
\item Camera or smartphone  ({\em black})
\item Photons ($\gamma$)  from the BLE 
\end{enumerate}

\noindent
Three of those world-lines ($b, c$, and $d$) must intersect (simultaneously) inside the same tiny spacetime volume ({\em hatched oval}). 
}
\hspace{0.5in}
\parbox{0.5in}{
\centering
\includegraphics[scale=0.35]{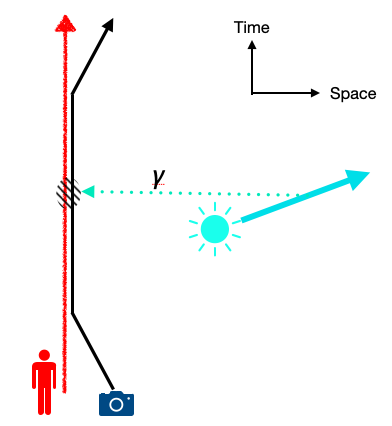}
}

\noindent
In other words, even if 
a smartphone or camera had been available to me during this BLE, the chance of all three world-lines intersecting is incredibly small. Thus, written documentation would most likely still constitute the best possible evidence.

In addition to the exceedingly small probability of capturing any kind of BLE image, 
there remains the equally difficult problem of establishing acceptance criteria. 
These criteria should include but not be limited to:
\begin{enumerate}
\item 
Detecting genuine images. 
A quick Internet search for ``ball lightning''  (of which there seem to be precious few meaningful results), 
can be categorized as known fakes or, capturing of undetermined refractive effects or, simply clever artwork acting as clickbait.   
None of the retrieved images or videos I found corresponded to anything like what has been described here. 
Regrettably, the situation seems to be not unlike that for claimed UFO sightings. 
\item 
Interpreting genuine images. This is not as simple as it may seem at first blush. 
For example, police {\em body-cams} are intended to be particularly helpful for clarifying otherwise chaotic and ambiguous dynamic circumstances. 
Often, however, the images in those videos raise more questions than they answer. 
Would a nighttime version of Fig.~\ref{fig:fireball} ``prove'' that it was a BLE? 
What {\em exactly} are you looking at?
\end{enumerate}

\noindent
Unlike spatially localized earthquake faults, 
even if we had cameras (and related detectors) densely integrated into existing electrical power-distribution networks, 
we still cannot define appropriate criteria for triggering BLE detection without creating an overwhelming number 
of false positives.

\section{Conclusion}   \label{sec:summary}
The BLE reported in Section~\ref{sec:event} was located at {\tt \small -37.795754, 145.074425}. 
The associated electrical storm officially started at 05:30:00 UTC  
near the central business district of Melbourne, Australia ({\tt \small -37.8200,144.9700}) on October 31, 1973. 
In terms of the attributes listed in Section~\ref{sec:intro}:
\begin{itemize}
\item {\bf Size:} Approximately that of a basketball (Fig.~\ref{fig:fireball}). 
\item {\bf Brightness:} Equivalent to vehicle headlights reflected from a wet surface. 
\item {\bf Lifetime:} Directly observed for a few seconds but possibly existed (unseen) for several minutes with 
the lightning-ball only spending about half that time between live power lines.
\item {\bf Motion:} Linear horizontal motion that appeared to have a constant speed of about 5 mph but might also have been slowly accelerating. 
\item {\bf Locomotion:}  Possibly driven by high-tension power lines for half its journey 
but then continued independently and linearly at a similar height until termination in the local park (Fig.~\ref{fig:path}). 
\item {\bf Electrical:}  Definitely associated with both severe atmospheric electrical conditions and 
proximate electrical power lines (Fig.~\ref{fig:wires}). The BLE also gave off sporadic sparks. 
Very localized damage to the terminus tree (Fig.~\ref{fig:trees}) was likely due to the lightning-ball exploding.  
\end{itemize}

\noindent
It is not absolutely certain where the BLE originated. 
It seemed to arise at the point where the lower wires attached to the corner power pole  in Fig.~\ref{fig:fireball}. 
Strictly speaking, however, that is the location where it first caught my attention. 
It is possible that it originated elsewhere  prior to my seeing it.

Being positioned inside a house, behind sealed windows, during a  noisy thunderstorm, no hissing sound or ozone smell 
was detected emanating from the BLE. 

Why did I not document this BLE in 1973?  If I did, it is now long lost. 
More likely, without some kind of physical evidence, making a timestamped note 
would have seemed no more convincing than the verbal description given to friends and acquaintances. 

A better course of action would have been to notify \href{https://twitter.com/BOM_Vic}{@BOM\_Vic} and, ideally, I should have done that: including reporting the damaged tree of Section~\ref{sec:tree}. 
However, even if I had thought of doing that, without any physical evidence, I would have quickly dismissed it   
under tha assumption that \href{https://twitter.com/BOM_Vic}{@BOM\_Vic} would simply have ignored it.  So, why bother?  
Furthermore,  it seems I was not alone in my self-defeatism. 
The incidents that led to this electrical storm being classified as {\em severe} in Section~\ref{sec:year} were not communicated by individuals. Rather, they appeared in  local newspaper reports  that were later scanned by \href{https://twitter.com/BOM_Vic}{@BOM\_Vic}. 
Moreover, unlike today, there was no Internet, no email, no web, no social media, and no arXiv.org to expedite either informal or formal communication.
Thus, {\em this} is the documentation.

The unanticipated irony is that during the subsequent decades, 
a considerable amount of supporting data has become available on the Internet, and 
applying those data to the BLE described here may actually have helped to enhance its scientific value.



\appendix
\section*{Appendix} \label{app:app}
\parbox[c]{2.25in}{
When I was eight years old, a neighbor presented me with 
his copy of an aging science book~\cite{book1949}. 
The hand-drawn illustration ({\em at right}) immediately caught my attention and became my first awareness of ball lightning. 
It looked so fantastic and frightening that I had to ask my father if it was real.   
I never imagined I would witness such an eerie and dangerous phenomenon. 

\medskip
Although the illustration is obviously an artist's conception that 
combines many different accounts into the same picture, it does show   
forked lightning and rain (cf. Section~\ref{sec:event}). 
The initial single ``fireball'' \href{https://en.wikipedia.org/wiki/Pair_production}{splits into a pair} 
that enter the dwelling via the chimney, rather than penetrating exterior walls  or windows.   
Once inside, they do penetrate walls and floors.  

\medskip
Unlike Fig.~\ref{fig:wires}, no external power-lines or internal AC wiring are depicted.  
The fireballs ``travel slowly'' but erratically---rather than linearly, as in Figs.~\ref{fig:fireball} and~\ref{fig:path}. 
One of them ({\em lower left}) explodes (cf. Section~\ref{sec:tree}) while  
the other ({\em lower right}) continues unchanged (after `killing five sheep'') and finally exits the doorway.
}
\hspace{0.1in} 
\parbox[c]{1.0in}{
\includegraphics[scale=0.25]{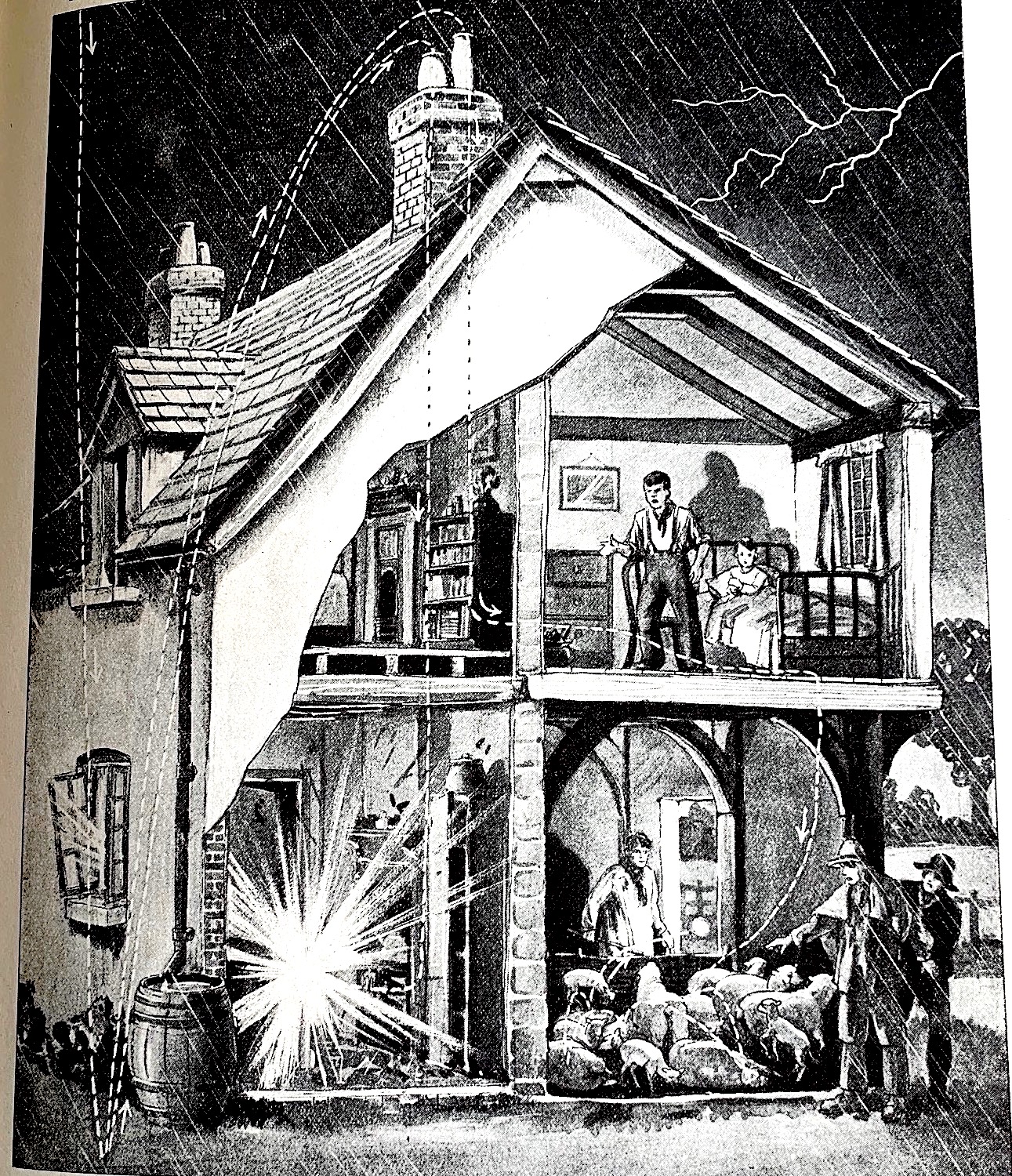}
}

Microwave-model simulations of ball lightning~\cite{wu2016} may account for many of these behaviors,   
including chimneys acting as a waveguide.
Consistent with other aspects of the illustration, but less fantastic, are the 1994 reports of similar BLE dynamics by many people in the town on Neuruppin, Germany~\cite{boerner2007}.


\begin{thebibliography}{99}
\setlength{\itemsep}{0pt}

\bibitem{wikiBL} 
``Ball lightning,'' 
\href{https://en.wikipedia.org/wiki/Ball\_lightning}{https://en.wikipedia.org/wiki/Ball\_lightning}, 
Last edited  Dec. 12 (2021)

\bibitem{book1949} 
{\em Wonders of Science Simplified: Three Volumes in One}, Metro Publications, New York, New York (1949)

\bibitem{boerner2007} 
D. B\"acker, H. Boerner, K. N\"ather, and S. N\"ather, 
``Multiple Ball Lightning Observations at Neuruppin,'' 
\href{https://www.researchgate.net/profile/Herbert-Boerner/publication/289306924_Multiple_ball_lightning_observations_at_Neuruppin_Germany/links/5cdad410a6fdccc9ddab8038/Multiple-ball-lightning-observations-at-Neuruppin-Germany.pdf}{The International Journal for Meteorology}, 
 32, 193 (2007)

\bibitem{QLD2011} 
S. Hughes,  ``Green fireballs and ball lightning,''  
\href{https://royalsocietypublishing.org/doi/pdf/10.1098/rspa.2010.0409}{Proc. Roy. Soc. A} 467, 1427–1448 (2011) 

\bibitem{keul2021} 
A. G. Keul,  
``A Brief History Of Ball Lightning Observations By Scientists And Trained Professionals,''  
\href{https://hgss.copernicus.org/articles/12/43/2021/}{Hist. Geo Space. Sci.}, 12, 43--56  (2021) 

\bibitem{medieval2022} 
G. Gasper and B. K. Tanner,  
``A Marvellous Sign and A Fiery Globe: A Medieval English Report of Ball Lightning,'' Weather (2022)


\bibitem{bbc} 
BBC World Service, 
\href{https://bbc.co.uk/programmes/w3ct1l41}{Science in Action}, 
Starting at 16:09, Aug. 13  (2021)    

\bibitem{survey} 
K. D. Stephan and R. Sonnenfeld, 
\href{https://docs.google.com/forms/d/e/1FAIpQLSc0Lxy_nmpR32C24BxPGxS0kc-QRKDNwB1oC_m0k_iIGBIJfA/viewform}{Ball Lightning  Eyewitness Report}, rev 1.07  (2021)

\bibitem{eos2021} 
J. Duncombe, ``Have You Seen Ball Lightning? Scientists Want to Know About It,''  
\href{https://eos.org/articles/have-you-seen-ball-lightning-scientists-want-to-know-about-it}{American Geophysical Union EOS}, 
 Jul. 15 (2021) 

\bibitem{nwsCG} 
National Weather Service, ``The Positive and Negative Side of Lightning,'' 
US Dept. Commerce,   
\href{https://www.weather.gov/jetstream/positive}{National Oceanic and Atmospheric Administration}, (online)

\bibitem{usaf2003} 
E. W., Davis, ``Ball Lightning Study,'' 
\href{https://documents2.theblackvault.com/documents/usaf/AFRL_2002-0039_Ball_Lightning_Study.pdf}{Final Report AFRL-PR-ED-TR-2002-0039}, Air Force Research Lab., Air Force Materiel Command, Edwards AFB, California  (2003)  \hspace{6pt} 

\bibitem{sturrock2016}
P. A. Sturrock, ``A Conjecture Concerning Ball Lightning,'' 
\href{https://arxiv.org/abs/1609.04238v1}{arXiv:1609.04238v1 [physics.gen-ph]}, 29 Aug (2016)


\bibitem{sturrock2017} 
P. A. Sturrock, ``\href{https://www.researchgate.net/publication/317753280_The_challenge_of_ball-lightning_Evidence_of_a_Parallel_Dimension}{The Challenge of Ball-Lightning: Evidence of a `Parallel Dimension'?},'' 
Journal of Scientific Exploration, Vol. 31, No. 1, pp. 84–91 (2017)


\bibitem{boerner2018} 
H. Boerner, 
``Analysis of Conditions Favorable for Ball Lightning Creation,'' 
\href{https://arxiv.org/abs/1606.04421v1}{arXiv:1606.04421v1 [physics.ao-ph]}  (2018)

\bibitem{funaro2018} 
D. Funaro, 
``A Model for Ball Lightning Derived from an Extension of the Electrodynamics Equations,'' 
in {\em Proc. VI Intl. Conf.  Atmosphere, Ionosphere, Safety}, Kaliningrad,
\href{https://arxiv.org/abs/1806.05555}{arXiv:1806.05555 [physics.gen-ph]}   (2018)

\bibitem{georgia2019} 
A. Abdallah, K. Castleberry, A. Spikes, K. Coumarbatch, J. Ballard-Myer, N. Palmer, and H. Mahabaduge, 
``A Review of Ball Lightning Models,'' 
\href{https://digitalcommons.gaacademy.org/cgi/viewcontent.cgi?article=1925&context=gjs}
{Georgia Journal of Science}, Vol. 77, No. 2, Article 18 (2019)


\bibitem{melpwr} 
T. Langstaff,  ``Electricity Networks,'' 
\href{https://www.ausnetservices.com.au/-/media/Files/AusNet/About-Us/Determining-Revenues/Distribution-Network/Customer-Forum/Week-1/Networks-101-Customer-Forum.ashx?la=en}{ AusNet Services}, Melbourne, Victoria, Australia  (2018)

\bibitem{bom} 
Bureau of Meteorology, Australia, 
\href{http://www.bom.gov.au/australia/stormarchive/}{Severe Storms Archive}  (online)


\bibitem{wu2016} 
H-C. Wu, ``Relativistic-Microwave Theory of Ball Lightning,'' 
\href{https://www.nature.com/articles/srep28263}{Scientific Reports}, 6, 28263 (2016)

\bibitem{Turner2003} 
D. J. Turner, 
"The Missing Science of Ball Lightning,"
Journal Scientific Exploration, Vol. 17, No. 3, pp. 435--496, 2003

\bibitem{Planck} 
``Ball Lightning,'' \href{https://www.ipp.mpg.de/2977926/kugelblitze}{IPP blog post}, 
Max Planck Institute for Plasma Physics (2021) 

\bibitem{spectral2014} 
J. Cen, P. Yuan, S. Xue,   ``Observation of the Optical and Spectral Characteristics of Ball Lightning,'' 
Physical Review Letters, 112 (3): 035001 (2014)


\end{thebibliography}
\end{document}